\documentclass[useAMS,usenatbib]{mn2e}

\usepackage{epsfig}
\usepackage{longtable}
\usepackage{times} 
\newcommand{\sw}{$Swift$}

\def \inte {\emph{INTEGRAL}}

\def \sw {{\it Swift}}
\def \rxte {\emph{RXTE}}
\def \src {\mbox{IGR~J18483$-$0311}}

\def \hcm {\hbox {\ifmmode $ atom cm$^{-2}\else atom cm$^{-2}$\fi}}

\def \ATel {Astron.\ Tel.}
\def \apj {ApJ}

\def \aap {A\&A}

\def \mnras {MNRAS}

\title[IGR~J18483$-$0311 with Swift]{\emph{Swift}/XRT monitoring of the Supergiant Fast X--ray Transient 
IGR~J18483$-$0311 for an entire orbital period}

\author[P.\ Romano et al.]{P.\ Romano$^{1}$, L.\ Sidoli$^{2}$, L.~Ducci$^{3,2}$, 
        G.\ Cusumano$^{1}$, V.\ La Parola$^{1}$, C.~Pagani$^{4}$, 
\newauthor K.L.~Page$^{5}$, J.A.~Kennea$^{4}$, D.N.~Burrows$^{4}$,  N.~Gehrels$^{6}$, V.~Sguera$^{7}$, A.~Bazzano$^{7}$ \\ 
$^{1}$INAF, Istituto di Astrofisica Spaziale e Fisica Cosmica,
        Via U.\ La Malfa 153, I-90146 Palermo, Italy\\
$^{2}$INAF, Istituto di Astrofisica Spaziale e Fisica Cosmica,
	Via E.\ Bassini 15,   I-20133 Milano,  Italy\\
$^{3}$Dipartimento di Fisica e Matematica, Universit\`a dell'Insubria, Via
      Valleggio 11, I-22100 Como, Italy \\
$^{4}$Department of Astronomy and Astrophysics, Pennsylvania State 
             University, University Park, PA 16802, USA\\
$^{5}$Department of Physics \& Astronomy, University of Leicester, LE1 7RH, UK\\
$^{6}$NASA/Goddard Space Flight Center, Greenbelt, MD 20771, USA\\
$^{7}$INAF, Istituto di Astrofisica Spaziale e Fisica Cosmica,
        Via Fosso del Cavaliere 100, I-00133, Roma, Italy \\
}

\begin{document}

\date{Accepted 2009 September 25. Received 2009 September 23; in original form 2009 July 23}
\pagerange{\pageref{firstpage}--\pageref{lastpage}} \pubyear{2009}

\maketitle

\label{firstpage}

\begin{abstract}
IGR~J18483$-$0311 is an X--ray pulsar with transient X--ray activity, 
belonging to the new class of High Mass X--ray Binaries called 
Supergiant Fast X--ray Transients. 
This system is one of the two members of this class, together with IGR~J11215--5952, 
where both the orbital (18.52~d) and spin period (21~s) are known.
We report on the first complete monitoring of the X--ray activity along an entire orbital 
period of a Supergiant Fast X--ray Transient. 
These \emph{Swift} observations, lasting 28~days, cover more than one entire orbital 
phase consecutively.   
They are a unique data-set, which allows us to constrain the different
mechanisms proposed to explain the nature of this new class of X--ray transients.
We applied the new clumpy wind model for blue supergiants developed by Ducci et al. (2009), 
to the observed X--ray light curve. Assuming an eccentricity of $e=0.4$,  
the X--ray emission from this source can be explained in terms of the accretion from a spherically 
symmetric clumpy wind, composed of clumps with different masses, ranging from $10^{18}$~g to 
5$ \times 10^{21}$~g. 
\end{abstract}



	\section{Introduction\label{igr18483:intro}}

The X--ray transient \src\ was discovered during the observations of the Galactic plane
with \inte\ in April 2003 \citep{Chernyakova2003}, 
when it reached a flux of 10~mCrab in the 15--40~keV energy range. 
Five more hard X--ray outbursts were reported by \citet{Sguera2007} with \inte, 
three of which exceeded one day in duration.
The strongest outburst ($\sim$1.8~days) occurred in April 2006 and reached a flux 
of 120~mCrab.
The \src\ broad band joint JEM-X/ISGRI \inte\ spectrum (3--50 keV) was fitted with an 
absorbed power law with a photon index, $\Gamma= 1.4\pm{0.3}$, a high absorption, 
$N_{\rm H} =9^{+5} _{-4}\times10^{22}$~cm$^{-2}$ 
(higher than the Galactic at $1.4\times10^{22}$~cm$^{-2}$), 
and a cutoff at $\sim22$~keV \citep{Sguera2007}. 

A periodicity at $18.55\pm{0.03}$~days was discovered 
in the ASM/\rxte\ light curves \citep{Levine2006:igr18483}, 
and was interpreted as the orbital period of a binary system.
\citet{Sguera2007} confirmed a similar period in the 20--40~keV ISGRI/\inte\ data 
($18.52\pm{0.01}$~days) and discovered pulsations at $21.0526\pm{0.0005}$~s 
with the X--ray monitor JEM-X. 
\citet{Skinner2008} refined the orbital period value to $18.518\pm0.005$\,d, 
by using the much denser sampling provided by the \sw/BAT data. 

From the observed values of the orbital and pulse periods, the position in the 
Corbet diagram  \citep{Corbet1986} suggested at first a Be/X--ray transient nature 
\citep{Sguera2007}, but the optical and infrared observations of the X--ray
error box estimated with \sw\ \citep{Sguera2007} revealed that the donor star is
a blue supergiant (B0.5Ia), and not a Be star, located at a distance of 3--4\,kpc 
\citep{Rahoui2008:18483}.
The X--ray position was later refined with $Chandra$ \citep{Giunta2009}, 
confirming this optical/IR association.
This  implied the identification of this source as a new member of 
the class of the Supergiant Fast X--ray Transients 
(SFXTs; \citealt{Sguera2005}, \citealt{Sguera2006}, \citealt{Negueruela2005a}, \citealt{Sidoli2009:cospar}),
although the dynamical range of its X--ray emission seems to be smaller than in other members of the same class.
\src\ and IGR~J11215--5952 are the only two SFXTs where both orbital and pulse periods have been discovered
(\citealt{Swank2007}, \citealt{SidoliPM2006}, \citealt{Romano2009:11215_2008}). 
%

 \begin{table}
 \begin{center}
 \caption{Summary of the {\it Swift}/XRT observations.\label{igr18483:tab:xrtobs} }
 \begin{tabular}{lllll}
 \hline
 \noalign{\smallskip}
 Seq.     & Start time  (UT)  & End time   (UT) & Exp. & Phase$^a$   \\ 
              &   &  &(s)        \\
  \noalign{\smallskip}
 \hline
 \noalign{\smallskip}
004	&  2009-06-11 14:42:17     &	   2009-06-11 21:13:56     &	   1606  &0.07   \\
005	&  2009-06-12 16:09:36     &	   2009-06-12 19:34:56     &	   1522  &0.12   \\
006	&  2009-06-13 18:08:11     &	   2009-06-13 21:26:57     &	   1774  &0.18   \\
007	&  2009-06-14 04:55:39     &	   2009-06-14 08:20:57     &	   1671  &0.21   \\
008	&  2009-06-15 10:03:12     &	   2009-06-15 23:02:58     &	   1977  &0.28   \\
009	&  2009-06-16 07:10:02     &	   2009-06-16 12:08:58     &       1857  &0.32   \\
010	&  2009-06-17 00:42:44     &	   2009-06-17 04:12:56     &       2047  &0.36   \\
012     &  2009-06-19 08:37:38     &	   2009-06-19 12:02:56     &	   2189  &0.49   \\
013     &  2009-06-20 13:32:09     &	   2009-06-20 16:57:56     &	   2291  &0.55   \\
016	&  2009-06-23 07:51:16     &      2009-06-23 12:53:56      &       2003  &0.70   \\
017	&  2009-06-24 09:32:31     &	   2009-06-24 14:35:56     &	   2711  &0.76   \\
018	&  2009-06-25 04:33:46     &	   2009-06-25 12:52:56     &	   1870  &0.81   \\
019	&  2009-06-26 15:43:27     &	   2009-06-26 19:06:57     &	   1884  &0.88   \\
020	&   2009-06-27 04:33:51    &	   2009-06-27 07:56:57     &	   1824  &0.91   \\
021	&   2009-06-28 03:16:17    &	   2009-06-28 08:14:56     &	   1922  &0.96   \\
022	&   2009-06-29 11:30:17    &	   2009-06-29 11:47:57     &	   1057  &0.03   \\
023	&   2009-06-30 06:47:17    &	   2009-06-30 11:54:57     &	   2616  &0.08   \\
024	&   2009-07-01 02:17:22    &	   2009-07-01 12:00:57     &	   2070  &0.13   \\
025	&   2009-07-02 10:22:32    &	   2009-07-02 13:23:57     &	   2408  &0.19   \\
026	&   2009-07-03 07:01:47    &	    2009-07-03 10:31:58    &	   2106  &0.24   \\
027	&   2009-07-04 07:05:17    &	    2009-07-04 10:36:27    &	   2560  &0.29   \\
029	&   2009-07-06 02:15:18     &	    2009-07-06 08:53:58    &	   1838  &0.39   \\
030	&   2009-07-08 07:32:02    &	    2009-07-08 07:37:56    &	    340  &0.51   \\
    \noalign{\smallskip} 
  \hline
  \end{tabular}
  \end{center}
  \begin{list}{}{} 
  \item[$^{\mathrm{a}}$ Calculated according to \citet{Sguera2007}.] 
  \end{list}   \end{table}

\begin{figure*}
\begin{center}
\vspace{-0.5truecm}
\includegraphics*[angle=270,width=18cm]{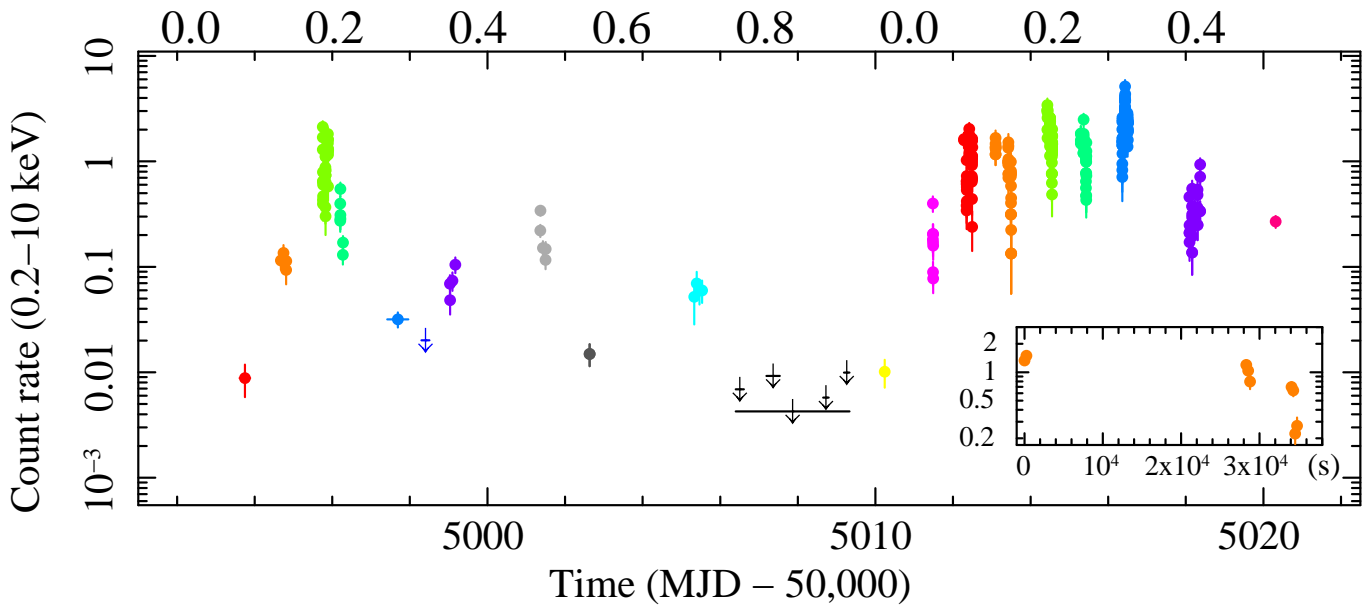}
\end{center}
\caption{\sw/XRT 0.2--10\,keV light curve of \src\ during our monitoring program, 
background-subtracted and corrected for pile-up, PSF losses, and vignetting. 
Downward-pointing arrows are 3$\sigma$ upper limits. 
The upper limit with the wide symbol centered on MJD 55007.9 ($\sim 0.004$ counts s$^{-1}$) 
is obtained accumulating the 4 observations (individual upper limits) between 
MJD 55006.5 and 55010.2. 
Different colours mark different 
observations (see Table~\ref{igr18483:tab:xrtobs}), with a colour scheme that 
generally mimics the phase (top axis) with a $P=18.52$ days \citep{Sguera2007}. 
The inset zooms on observation 024.
}
\label{igr18483:fig:xrtlcv}
\end{figure*}

 	 \section{Observations and Data Reduction\label{sax1818:dataredu}}

The observations of \src\ were obtained as a Target of Opportunity (ToO) monitoring program
with \sw. As shown in Table~\ref{igr18483:tab:xrtobs}, the ToO observations started on
2009 June 11 with $\sim2$\,ks per day. 
The campaign lasted 28 days divided in 23 observations for a total
on-source exposure of $\sim44$\,ks.

The XRT data were processed with standard procedures 
({\sc xrtpipeline} v0.12.1), filtering and screening criteria by using 
{\sc FTOOLS} in the {\sc Heasoft} package (v.6.6.1).  
Both WT and PC events were considered. 
The selection of event grades was 0--2 and 0--12, 
for WT and PC data, respectively (\citealt{Burrows2005:XRTmn}). 
We corrected for pile-up when required. 
The light curves were also corrected for PSF losses, vignetting and 
background-subtracted.
Ancillary response files were generated with {\sc xrtmkarf},
to account for different extraction regions, vignetting, and
PSF corrections. We used the spectral redistribution matrices v011 in CALDB.

The UVOT observed the target simultaneously with the XRT with the $v$ filter
(observations 004 through 006), and with the $u$ filter 
(observations 007 though 010).
For the remainder of the campaign \src\ was observed with the 
`Filter of the Day', i.e.\ the filter chosen for all observations 
to be carried out during a specific day in order to minimize the filter 
wheel usage \citep{Roming2005:UVOTmn}. 
The data analysis was performed using the {\sc uvotimsum} and 
{\sc uvotsource} tasks included in the {\sc FTOOLS}. The latter 
task calculates the magnitude through aperture photometry within
a circular region and applies specific corrections due to the detector
characteristics. The reported magnitudes are on the UVOT photometric 
system described in \citet{Poole2008:UVOTmn}, and 
are not corrected for Galactic extinction. 
At the position of \src, no detection was achieved down to a limit of 
$v>21.07$ mag and $u>21.19$ mag. 

All quoted uncertainties are given at 90\,\% confidence level for 
one interesting parameter unless otherwise stated. 
The spectral indices are parameterized as $F_{\nu} \propto \nu^{-\alpha}$, 
where $F_{\nu}$ (erg cm$^{-2}$ s$^{-1}$ Hz$^{-1}$) is the 
flux density as a function of frequency $\nu$; 
we adopt $\Gamma = \alpha +1$ as the photon index, 
$N(E) \propto E^{-\Gamma}$ (ph cm$^{-2}$ s$^{-1}$ keV$^{-1}$).

 \begin{table*} 	
 \begin{center} 	
 \caption{XRT spectroscopy. \label{igr18483:tab:xrtspec}} 	
 \label{} 	
 \begin{tabular}{lrlllrlllll} 
 \hline 
 \noalign{\smallskip} 
Seq.           & Power-law   &           &                     &                        & Black body & & & & & \\
            &$N_{\rm H}$  &$\Gamma$  &Flux$^a$ &$\chi^{2}_{\rm red}/dof$ &$N_{\rm H}$  &$kT$  &$R_{\rm BB}$  &Flux$^{a}$ &L$^{b}$ & $\chi^{2}_{\rm red}/dof$  \\
\noalign{\smallskip} 
           &   (10$^{22}$~cm$^{-2}$)           &           & (2--10\,keV)    & C-stat (\%)$^{c}$ & (10$^{22}$~cm$^{-2}$) &   (keV)   &   (km)$^d$  & (2--10\,keV)  & (2--10\,keV)   & C-stat (\%)$^{c}$ \\
 \hline 
005&$6.0_{-2.5}^{+3.4}$  &$1.4_{-0.8}^{+0.9}$ &$1.90_{-0.94}^{+0.37}$  &$102.1(48.3)$ &$3.5_{-1.6}^{+2.2}$  &$1.6_{-0.4}^{+0.6}$  &$0.14_{-0.01}^{+0.03}$&$1.46_{-0.83}^{+0.15}$  &$0.2$ &$100.9(31.90)$ \\
006&$6.3_{-1.2}^{+1.5}$  &$1.7_{-0.4}^{+0.4}$ &$15.46_{-3.45}^{+0.54}$ &$1.1/34$      &$3.1_{-0.7}^{+0.8}$  &$1.6_{-0.2}^{+0.2}$  &$0.43_{-0.14}^{+0.22}$&$11.3_{-1.99}^{+0.79}$  &$1.2$ &$1.1/34$ \\
007&$4.8_{-2.1}^{+3.8}$  &$1.3_{-0.7}^{+0.9}$ &$2.72_{-0.92}^{+0.39}$  &$1.2/9$       &$2.5_{-1.2}^{+2.0}$  &$1.6_{-0.3}^{+0.5}$  &$0.17_{-0.02}^{+0.04}$&$2.06_{-0.99}^{+0.24}$  &$0.2$ &$0.9/9$ \\
010&$6.6_{-2.2}^{+2.7}$  &$2.1_{-0.7}^{+0.8}$ &$1.45_{-0.68}^{+0.12}$  &$119.0(27.2)$ &$3.5_{-1.4}^{+1.7}$  &$1.3_{-0.2}^{+0.3}$  &$0.18_{-0.02}^{+0.05}$&$1.00_{-0.46}^{+0.10}$  &$0.1$ &$117.5(10.58)$ \\
012&$6.4_{-2.1}^{+3.0}$  &$1.8_{-0.6}^{+0.7}$ &$2.78_{-1.05}^{+0.19}$  &$1.2/13$      &$3.2_{-1.2}^{+1.7}$  &$1.5_{-0.2}^{+0.3}$  &$0.20_{-0.02}^{+0.05}$&$1.98_{-0.71}^{+0.16}$  &$0.2$ &$1.1/13$ \\
022&$16.5_{-6.1}^{+7.7}$ &$2.3_{-1.1}^{+1.2}$ &$5.79_{-4.33}^{+0.30}$  &$97.43(26.7)$ &$10.7_{-4.0}^{+5.2}$ &$1.5_{-0.3}^{+0.6}$  &$0.27_{-0.08}^{+0.28}$&$3.43_{-2.24}^{+0.22}$  &$0.4$ &$98.02(12.17)$ \\
023&$6.6_{-1.2}^{+1.4}$  &$1.4_{-0.3}^{+0.3}$ &$17.44_{-3.38}^{+0.51}$ &$0.9/51$      &$3.7_{-0.7}^{+0.8}$  &$1.7_{-0.2}^{+0.2}$  &$0.39_{-0.10}^{+0.15}$&$13.26_{-2.14}^{+0.62}$ &$1.4$ &$1.0/51$ \\
024&$10.4_{-2.3}^{+2.9}$ &$1.5_{-0.4}^{+0.5}$ &$19.97_{-6.79}^{+0.58}$ &$0.7/25$      &$6.3_{-1.4}^{+1.8}$  &$1.9_{-0.3}^{+0.3}$  &$0.37_{-0.11}^{+0.19}$&$14.78_{-4.13}^{+0.84}$ &$1.6$ &$0.7/25$ \\
025&$7.1_{-1.0}^{+1.1}$  &$1.6_{-0.3}^{+0.3}$ &$32.83_{-4.92}^{+0.87}$ &$1.1/71$      &$3.8_{-0.6}^{+0.7}$  &$1.7_{-0.1}^{+0.2}$  &$0.56_{-0.27}^{+0.38}$&$24.23_{-3.19}^{+0.92}$ &$2.6$ &$1.2/71$  \\
026&$6.3_{-1.1}^{+1.3}$  &$1.2_{-0.3}^{+0.3}$ &$20.83_{-3.87}^{+0.88}$ &$1.0/42$      &$3.7_{-0.7}^{+0.8}$  &$1.9_{-0.2}^{+0.2}$  &$0.38_{-0.09}^{+0.14}$&$16.68_{-3.08}^{+0.94}$ &$1.8$ &$1.0/42$  \\
027&$6.6_{-0.8}^{+0.9}$  &$1.3_{-0.2}^{+0.2}$ &$43.22_{-5.46}^{+1.18}$ &$0.8/81$      &$3.8_{-0.5}^{+0.6}$  &$1.9_{-0.1}^{+0.2}$  &$0.56_{-0.22}^{+0.29}$&$34.14_{-3.64}^{+1.39}$ &$3.7$ &$0.8/81$ \\
029&$7.9_{-1.9}^{+2.6}$  &$1.8_{-0.5}^{+0.5}$ &$6.55_{-2.17}^{+0.20}$  &$1.4/25$      &$4.3_{-1.1}^{+1.4}$  &$1.6_{-0.2}^{+0.2}$  &$0.28_{-0.05}^{+0.09}$&$4.64_{-1.18}^{+0.30}$  &$0.5$ &$1.3/25$ \\
 \hline 
high$^{\mathrm{e}}$   &$5.4_{-0.4}^{+0.5}$ &$1.0_{-0.1}^{+0.1}$ &$25.77_{-1.48}^{+0.46}$ &$1.0/222$ &$3.2_{-0.3}^{+0.3}$ &$2.1_{-0.1}^{+0.1}$ &$0.37_{-0.04}^{+0.05}$  &$21.25_{-1.21}^{+0.56}$ &$2.3$ &$1.2/222$ \\
medium$^{\mathrm{e}}$ &$6.4_{-0.5}^{+0.6}$ &$1.2_{-0.1}^{+0.1}$ &$16.47_{-1.06}^{+0.28}$ &$1.0/212$ &$3.7_{-0.3}^{+0.4}$ &$2.0_{-0.1}^{+0.1}$ &$0.32_{-0.03}^{+0.03}$  &$13.14_{-0.69}^{+0.33}$ &$1.4$ &$1.1/212$ \\
low$^{\mathrm{e}}$    &$6.7_{-0.8}^{+0.9}$ &$1.5_{-0.2}^{+0.2}$ &$1.82_{-0.06}^{+0.16}$  &$1.0/115$ &$3.6_{-0.5}^{+0.5}$ &$1.7_{-0.1}^{+0.1}$ &$0.130_{-0.002}^{+0.003}$  &$1.38_{-0.11}^{+0.04}$ &$0.2$ &$0.8/115$ \\
  \noalign{\smallskip}
  \hline
  \end{tabular}
  \end{center}
  \begin{list}{}{} 
  \item[$^{\mathrm{a}}$ Fluxes (corrected for the absorption) are in units of $10^{-11}$ erg~cm$^{-2}$~s$^{-1}$.] 
  \item[$^{\mathrm{b}}$ Luminosities in units of $10^{35}$ erg~s$^{-1}$, assuming a distance of 3~kpc.] 
  \item[$^{\mathrm{c}}$ Cash statistics (C-stat) and percentage of $10^4$ Monte Carlo realizations that
had statistics $<$ C-stat.] 
  \item[$^{\mathrm{d}}$ Blackbody radii are in units of km, assuming the optical counterpart distance of 3~kpc.] 
  \item[$^{\mathrm{e}}$ Intensity-selected spectra. High corresponds to  
$CR>1$ counts s$^{-1}$, medium to $0.5<CR<1$ counts s$^{-1}$, and 
low to $CR<0.5$ counts s$^{-1}$.]

  \end{list}
  \end{table*}

 	 \section{Results \label{igr18483:results} }

 	 \subsection{Light curve  \label{igr18483:lcvs} }

Fig.~\ref{igr18483:fig:xrtlcv} shows the 0.2--10\,keV light curve of \src\ 
of the whole campaign after background-subtraction and  pile-up, PSF losses, and 
vignetting corrections. Each bin contains a minimum of 20 source counts.  
The light curve starts at phase 0.07, assuming a period of 18.52\,d and an initial epoch 
MJD 53844.2  \citep{Sguera2007}, and monitors the flux state through a whole period
until the following phase 0.51. Superimposed on the long-term orbital modulation, 
variability is seen on short time scales, as shown in the inset of 
Fig.~\ref{igr18483:fig:xrtlcv}, where a variation by factor of 5.3 
in count rate is observed to occur in $\sim 1.7$ hr. 
This behaviour has been observed in several SFXTs  
\citep[e.g.,][]{Sidoli2008:sfxts_paperI,Romano2009:sfxts_paperV}.

The lowest point in the campaign is a 3$\sigma$ upper limit 
reached on MJD 55006.5--55010.2 at 0.004 counts s$^{-1}$ 
(combined observations 017--020, total on-source exposure of 8.3\,ks),
and corresponds to an observed (unabsorbed) flux of $3.9\times10^{-13}$ 
($2.1\times10^{-12}$) erg cm$^{-2}$ s$^{-1}$, 
if we assume the \emph{XMM--Newton} spectrum reported by 
\citet[][photon index $\Gamma=2.5$, absorbing column 
$N_{\rm H}=7.7\times 10^{22}$ cm$^{-2}$]{Giunta2009}. 
The corresponding luminosity is $2.3\times10^{33}$ erg s$^{-1}$ 
(assuming the optical counterpart distance of 3~kpc); 
to date this is the lowest quiescent X-ray flux value reported in the literature
for this source. 
The peak count rate is reached on MJD 55016.4 at $\ga 5$ counts s$^{-1}$; 
therefore, the observed dynamical range of this source is at least 1200.

In order to search for spin periodicity, the arrival times of all
selected events have been converted to the Solar System barycentric
frame, using the {\sc barycor} code.
The $Z^2$ test \citep{Buccheri1983} on the fundamental harmonics
was applied to a sample of source photon arrival times for each observation.
The search for a timing feature was performed within the frequency interval 
0.01--0.19\,s$^{-1}$. No presence of coherent pulsations was detected.

\begin{figure}
\begin{center}
\includegraphics*[angle=0,scale=0.40]{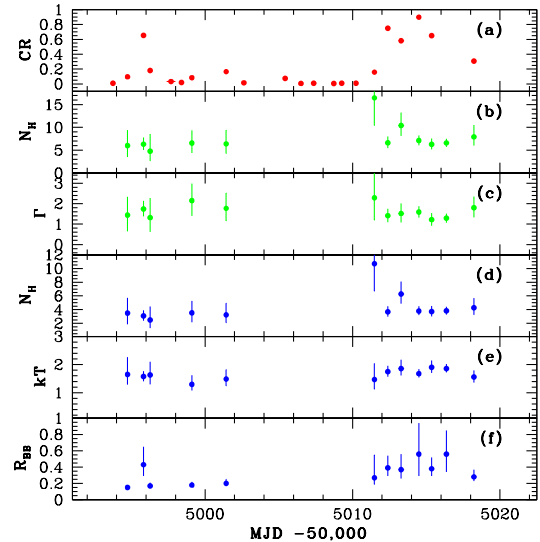}
\end{center}
\vspace{-0.5truecm}
\caption{Spectral parameters as a function of time (see Table~\ref{igr18483:tab:xrtspec}).
{\bf (a)} \sw/XRT light curve in the 0.2--10\,keV energy band at a day resolution;  
Spectral parameters of the absorbed power-law fit, 
$N_{\rm H}$, and photon index $\Gamma$ [{\bf (b)} and {\bf (c)}]; 
 Spectral parameters of the absorbed blackbody fit, 
$N_{\rm H}$, temperature $kT$, and blackbody radius [{\bf (d)}, {\bf (e)}, and {\bf (f)}].
}
\label{igr18483:fig:spectrallcvs}
\end{figure}

 	 \subsection{Spectra \label{igr18483:spectra} }

Spectra were extracted for each segment in which a detection was obtained 
and a minimum of 120 source counts were available. The data were rebinned with 
at least 20 counts bin$^{-1}$ to allow $\chi^2$ fitting, except when the 
statistics were poor, in which case we adopted \citet{Cash1979} statistics and 
data binned to 1 count bin$^{-1}$, instead. 
The simple models we considered were absorbed power laws and absorbed blackbodies. 
The fits were performed in the 0.3--10\,keV energy band. 
The results are reported in Table~\ref{igr18483:tab:xrtspec}, while  
the spectral parameters as a function of time are shown in 
Fig.~\ref{igr18483:fig:spectrallcvs}. 
In particular, for the spectrum of observation 006, the highest in flux during the
first peak ($\sim 1.5\times 10^{-10}$ erg cm$^{-2}$ s$^{-1}$), 
an absorbed power-law model yielded a high absorbing column  
$N_{\rm H}=(6.3_{-1.2}^{+1.5})\times 10^{22}$ cm$^{-2}$, and 
a photon index $\Gamma=1.7\pm0.4$ (see Table~\ref{igr18483:tab:xrtspec}). 
As a comparison, observation 025, roughly at the same phase, 
yielded $N_{\rm H}=(7.1_{-1.0}^{+1.1})\times 10^{22}$ cm$^{-2}$, and $\Gamma=1.6\pm0.3$,
hence consistent with observation 006. 
Despite the large observed variations in flux throughout the campaign, 
the spectral parameters do not vary significantly within the large uncertainties, 
with the exception of the blackbody radii. 

\begin{figure}
\begin{center}
\centerline{\includegraphics[width=6cm,angle=270]{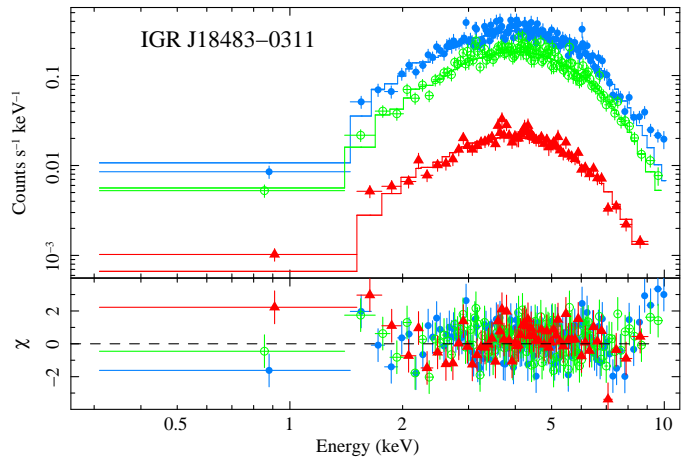}}
\caption[XRT spectra and residuals]{  Intensity-selected spectroscopy. 
            Upper panel: \sw/XRT data fit with an absorbed power law. 
            Lower panel: the residuals of the fit (in units of standard deviations). 
            Filled blue circles, green empty circles, and red filled triangles
            mark high, medium, and low state, respectively. 
                 }
	\label{igr18483:fig:specfits}   
       \end{center}
       \end{figure}

To further investigate the spectral properties of the sources in several states, 
we accumulated all events collected during the current campaign. 
We extracted events within three intensity levels depending on count rate, namely,
$CR>1$ counts s$^{-1}$ (high, 5447 counts), $0.5<CR<1$ counts s$^{-1}$ (medium, 5264 counts), and 
$CR<0.5$ counts s$^{-1}$ (low, 2670 counts).  
We created exposure maps for each of these intensity-selected event files  
and then combined them (and their exposure maps),  
and extracted a single spectrum for state. 
The generation of ancillary response files and spectral fitting 
were performed in the same fashion as for the single observations. 
The fit results are reported in Table~\ref{igr18483:tab:xrtspec}
and the spectra are shown in Fig.~\ref{igr18483:fig:specfits}. 
Even with the higher statistics afforded by accumulating all events in 
three intensity states, no significant variations in the column density
could be derived. We can confirm, however, that the $N_{\rm H}$ is always 
in excess of the Galactic one, $1.4\times 10^{22}$ cm$^{-2}$, 
consistently with that found by \citet{Sguera2007}. 
Similarly to what was found in a sample of 4 SFXTs \citep{Romano2009:sfxts_paperV},
our fits indicate either a hard power law or hot black body. 
We also note that all spectral fits with an absorbed black body resulted in 
radii of the emitting black body region of only a few hundred meters 
(see Table~\ref{igr18483:tab:xrtspec}), consistent with being emitted from a
small portion of the neutron star surface, such as its polar caps 
\citep[see, ][]{Romano2009:sfxts_paperV}.

 	 \section{Discussion \label{igr18483:discussion}}

\begin{figure}
\vspace{-0.5cm}
\begin{center}
\includegraphics[width=9cm]{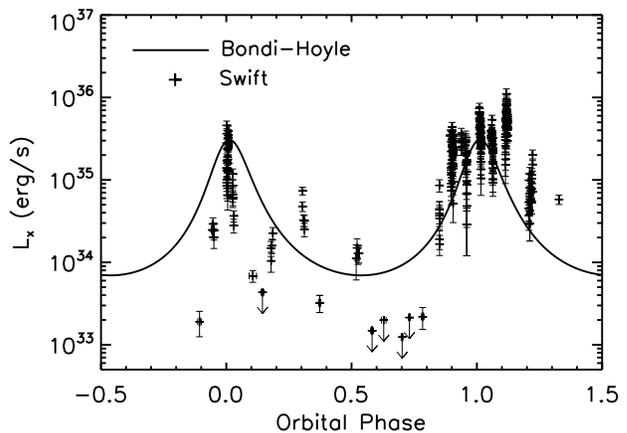}
\end{center}
\vspace{-0.2cm}
\caption{Comparison of the \emph{Swift}/XRT light curve of \src\  (crosses)
with the prediction of Bondi-Hoyle accretion from a spherically symmetric and homogeneous wind. 
We assumed a distance of 3\,kpc. 
The model-dependent orbital phase $\phi=0$ corresponds to 54995.83~MJD.}
\label{igr18483:fig:bondihoyle}
\end{figure}

In this paper we report on the first complete
monitoring of the X--ray activity along an entire orbital period of a 
Supergiant Fast X--ray Transient.
This makes these \sw\ observations a unique data-set, which allows us to constrain the different
mechanisms proposed to explain the nature of this new class of X--ray transients.
\src\ and IGR~J11215--5952 are the only SFXTs where both the orbital and spin periods
are known, although the two systems are very different  
(P$_{\rm orb}\sim$165~days, P$_{\rm spin}\sim$190~s, in IGR~J11215--5952; 
see, e.g., \citealt{Romano2009:11215_2008}).
 
The \sw\ light curve appears to be highly modulated, with two maxima, separated by a time
interval consistent with the orbital period of $\sim$18.5~days. 
A lower limit of 1200 to the dynamical range can be obtained from the observed light curve.
The different duration of the two outburst peaks monitored with \sw\ 
is probably the result of both a different sampling and a high intrinsic X--ray variability.
The second peak has a duration of several days, as previously observed by INTEGRAL \citep{Sguera2007}.

The modulation of the overall shape of the light curve with the orbital phase 
can be interpreted as wind accretion along a highly eccentric orbit.
Thus, we  applied different models for the wind accretion to gain information on the source parameters.

\begin{figure}
\vspace{-0.5cm}
\begin{center}
\includegraphics[width=9cm]{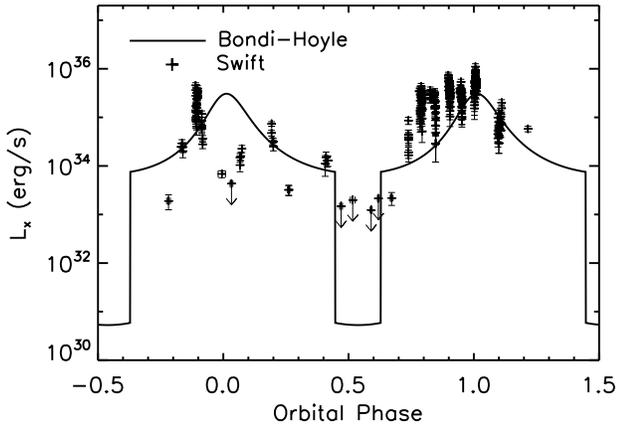}
\end{center}
\vspace{-0.2cm}
\caption{Comparison of the \emph{Swift}/XRT light curve of \src\  (crosses)
with the prediction of Bondi-Hoyle accretion from a spherically symmetric and homogeneous wind,  
assuming a magnetic field $B=4 \times 10^{11}$~G. 
The model-dependent orbital phase $\phi=0$ corresponds to 54997.69~MJD.
}
\label{igr18483:fig:centrif_inib}
\end{figure}

The simplest case is a Bondi-Hoyle accretion from a 
spherically symmetric and homogeneous wind.
Assuming a supergiant mass of $M_{\rm OB}=33\,M_{\odot}$
and a radius, $R_{\rm OB}=33.8\,R_{\odot}$ \citep{Searle2008}, together with
the known orbital period of $P_{\rm orb}=18.52$~d,
we tried to account for the overall shape of the X--ray light curve, 
leaving the orbital eccentricity $e$ as a free parameter.
We assumed the following values for the wind properties:
a terminal velocity $v_{\infty}$ in the range $1400$--$1800$~km~s$^{-1}$, 
a mass loss rate $\dot{M}$ in the range $0.4$--$1.5  \times 10^{-6}$~M$_{\odot}$~yr$^{-1}$
[see \citet{Searle2008}, \citet{Lefever2007}]. 
In this framework, we obtain the best agreement with the observed light curve by assuming 
an eccentricity $e=0.4$, 
$v_{\infty}=1800$~km~s$^{-1}$, $\dot{M}=5 \times 10^{-7}$~M$_{\odot}$~yr$^{-1}$,
$\beta=1$.
Fig.~\ref{igr18483:fig:bondihoyle} shows the comparison of the 
model predictions and the observed \sw/XRT light curve  
(in units of ergs s$^{-1}$, by assuming a distance of 3\,kpc). 
The model roughly reproduces the shape of the X--ray light curve due to the 
orbital modulation, with the largest deviation from the observations being 
in the time interval MJD 55006.5--55010.2, where we observed 
4 upper limits.
Note, however, that we cannot be sure that
the low intensity extends for 4 days continuously, because the four 
\sw\ observations consist of short snapshots.
We investigated the possibility that these upper limits
could be due to the onset of a 
centrifugal inhibition for the accretion \citep{Davidson1973}.
For the above adopted set of wind and orbital parameters, 
we calculated a new X--ray light curve, finding that 
a low magnetic field of the neutron star of 
$B\approx 4 \times 10^{11}$~G  could be responsible for the upper limits
at MJD 55006.5--55010.2 (see Fig.~\ref{igr18483:fig:centrif_inib}).

An alternative scenario to explain the low intensity state is 
an X--ray eclipse of the neutron star by the supergiant companion.
Assuming a circular orbit, 
we obtain a lower limit to the radius of the supergiant
star \citep{Rappaport1983} of $R_{\rm OB}=46$~$R_{\odot}$, which
is too large for a B0.5a supergiant \citep{Searle2008}.
On the other hand, our modelling of the X--ray curve suggests a high eccentricity
of at least $e=0.4$. Adopting this eccentricity, we
derived an expected value for the supergiant radius of $R_{\rm OB}=39.5~R_{\odot}$.
The radii of B0.5Ia stars are usually smaller than this value, but
there are several exceptions with $R \approx 40$~$R_{\odot}$ \citep{Searle2008,Lefever2007}. 
Therefore we cannot exclude that an eclipse is responsible for the low luminosity state,
in an eccentric orbit.

Although both centrifugal inhibition and an eclipse can 
reconcile the observed low intensity state with the Bondi-Hoyle 
accretion predictions, it is also clear that the 
spherically symmetric and homogeneous wind only reproduces the 
overall shape of the X--ray light curve. It cannot, indeed,  
account for the very large spread around the average behaviour due to the 
orbital modulation and, most of all, the remakable 
short time scale variability (see, for example, the inset 
in Fig.~\ref{igr18483:fig:xrtlcv}, where a 
variation by a factor of 5.3 in count rate is observed 
in $\sim 1.7$ hours). 
The observed short time scale variability can be naturally explained 
by the accretion of single clumps composing the donor wind. 
Thus, in order to improve the agreement between the observed
and the calculated light curve, we applied 
the isotropic clumpy wind model proposed by \citet{Ducci2009}.

The \citet{Ducci2009} model was developed to investigate the effects 
of accretion from a clumpy wind on the luminosity and variability 
properties of HMXBs. 
It assumes that a fraction of the stellar wind is in the 
form of clumps with a power law mass distribution
\begin{equation} \label{Npunto}
p(M_{\rm cl})=k \left ( \frac{M_{\rm cl}}{M_{\rm a}} \right)^{-\zeta}
\end{equation}
in the mass range $M_{\rm a}$--$M_{\rm b}$.
The rate of clumps produced by the supergiant is related to the
total mass loss rate $\dot{M}_{\rm tot}$ by 
\begin{math} 
\label{Ndot emitted}
\dot{N}_{\rm cl} = \frac{f \dot{M}_{\rm tot}}{<M>} \mbox{ \ \
clumps s}^{-1},
\end{math}
where $f=\dot{M}_{\rm cl} / \dot{M}_{\rm tot}$ is the fraction of
mass lost in  clumps  and $<M>$ is the average clump mass, which can
be computed from Eq.~(\ref{Npunto}).
Clumps are driven radially outward by absorption of UV spectral
lines. The following clump velocity profile is assumed: 
\begin{math} \label{legge_velocita}
v_{\rm cl}(r) = v_{\infty}\left (1 - 0.9983\frac{R_{\rm OB}}{r}
\right )^{\beta}, 
\end{math}
where $v_{\infty}$ is the terminal wind velocity, $R_{\rm OB}$ is the
radius of the supergiant, $0.9983$ is a parameter which ensures
that $v(R_{\rm OB}) \approx 10$~km~s$^{-1}$, and $\beta$ is a
constant in the range $\sim$$0.5$--$1.5$. 
The model further assumes that the clumps are confined by ram 
pressure of the ambient gas. 
By exploring different distributions for the clump masses and initial 
dimensions, 
the model can be used to compute the expected X-ray light
curves in the framework of the Bondi--Hoyle accretion theory, 
modified to take into account the presence of clumps.
We sought the set of wind parameters 
yielding the best agreement between the calculated and the observed light curve.
We found that the observed light curve is reproduced well by this wind model by 
assuming the following parameter values:  
an eccentricity $e=0.4$, 
a mass loss rate $\dot{M}_{\rm tot}=2 \times 10^{-7}$~M$_{\odot}$~yr$^{-1}$,
$v_{\infty}=1800$~km~s$^{-1}$,
$\beta=1$,
a fraction of mass lost in clumps $f=0.75$,
a mass distribution power law index $\zeta=1.1$,
a power law index of the initial clump dimension distribution $\gamma=-1$
(where $\dot{N}_{M_{cl}} \propto R_{\rm cl}^{\gamma} \mbox{ \ \ clumps s}^{-1}$),  
a minimum clump mass $M_{\rm a}=10^{18}$~g
and a maximum clump mass $M_{\rm b}=5 \times 10^{21}$~g; 
moreover, we adopt the force multiplier parameter obtained 
by \citet{Shimada1994} for a B0.5Ia star
($k = 0.375$, $\alpha = 0.522$, $\delta = 0.099$). 

\begin{figure}
\begin{center}

\vspace{-1.cm}
\includegraphics[width=9cm]{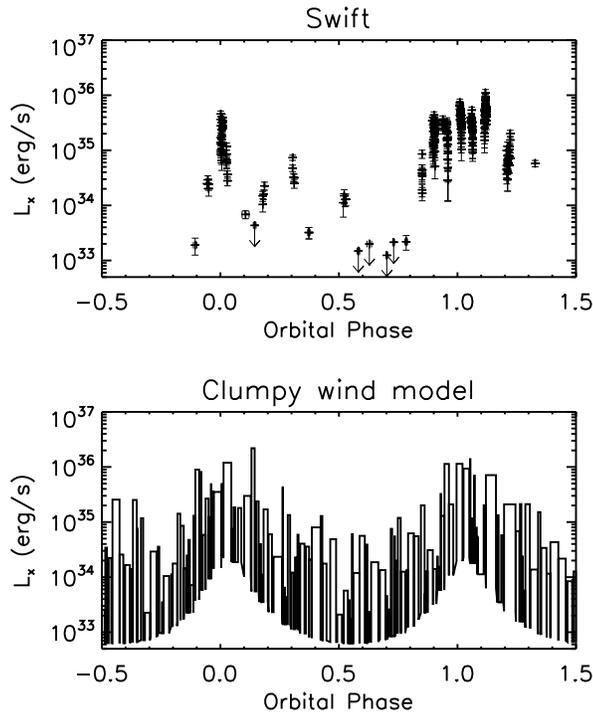}
\end{center}

\vspace{-1.5cm}
\caption{Comparison of the \emph{Swift}/XRT light curve of \src\  (top)
with the prediction (bottom) of the new clumpy wind model of \citet{Ducci2009}.
The model-dependent orbital phase $\phi=0$ corresponds to 54995.83~MJD.
}
\label{igr18483:fig:clumpymodel}
\end{figure}

\begin{figure*}[t]
\includegraphics*[angle=0,width=18cm]{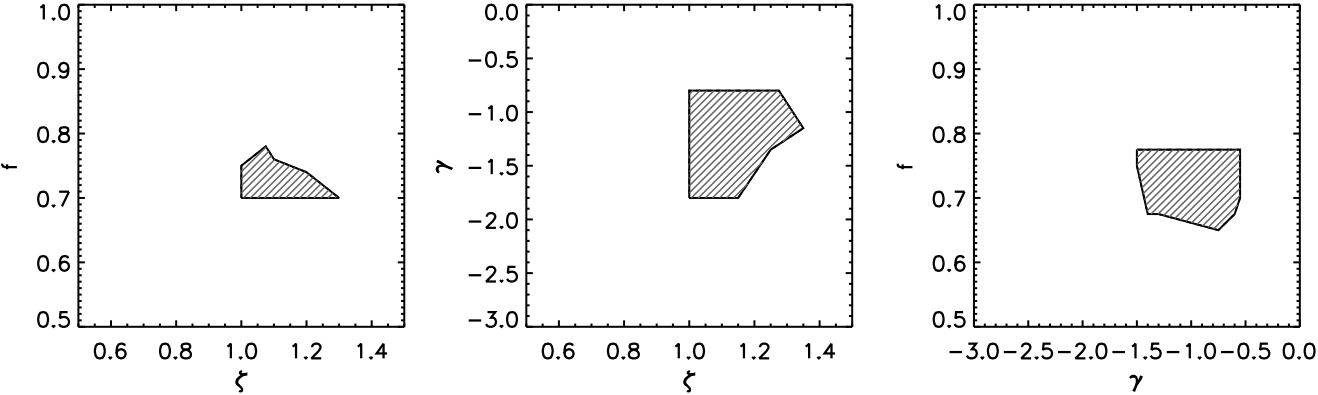}
\vspace{-0.5cm}

\caption{ 
Graphs of the allowed parameters $\zeta$, $f$, $\gamma$ (filled region),
         obtained from the comparison between the observed and 
         the calculated light curves.
}
\label{igr18483:fig:allowed-ranges}
\end{figure*}

Fig.~\ref{igr18483:fig:clumpymodel} shows the comparison of the \emph{Swift}/XRT 
light curve of \src\ with the isotropic clumpy wind model prediction.
Further acceptable solutions can be found by assuming 
wind parameters $\zeta$, $f$, $\gamma$ in the allowed 
ranges plotted in Fig.~\ref{igr18483:fig:allowed-ranges},
and $e=0.4 \pm 0.1$,
$\dot{M}_{\rm tot}=(2\pm 1) \times 10^{-7}$~M$_{\odot}$~yr$^{-1}$,
$v_\infty=1800$~km~s$^{-1}$, $\beta=1$, 
$10^{18} \leq M_{cl} \leq 10^{21}$~g.
As Fig.~\ref{igr18483:fig:allowed-ranges} demonstrates, the comparison of the
observed light curve with the clumpy wind model
allowed us to constrain the parameters responsible for the degree
of inhomogeneity of the wind.
In particular, we found that a very large fraction of the mass lost from 
the supergiant is contained in the clumps ($0.7 \la f \la 
0.78$),
and we obtain the value of $\zeta$ (which controls the shape
of the clump formation rate distribution) with an accuracy of $\sim$15\%.

Fig.~\ref{igr18483:fig:clumpymodel} shows that  the peak luminosities, the 
dynamic range involved by the flares, and 
the orbital modulation and the low luminosity state  (MJD 55006.5--55010.2)
observed are reproduced well by the clumpy wind model, 
even without invoking either a centrifugal barrier or an X--ray eclipse.
Indeed, from the calculated light curve, 
we determined that the probability to observe
the source at the inter-clump luminosity level 
in the range of phase $0.2<\phi<0.8$ is $\sim 25$\%:
with the binomial distribution function, we obtain a
probability to measure 4 low luminosity states of $\sim 5$\%.
Therefore, the upper limits can be explained with the accretion of the 
intra-clump wind with a low density,  even 
without invoking centrifugal inhibition or an eclipse.    
Finally,  we note that the wind parameters we obtain applying our spherical clumpy 
wind model \citep{Ducci2009} are
very similar to those explaining the Vela X--1 X--ray light curve. 
Indeed, the two systems have very similar donor stars.


\section*{Acknowledgments}

We thank the {\it Swift} team duty scientists and science planners. 
We also thank the remainder of the {\it Swift} XRT and BAT teams,
S.\ Barthelmy and J.A. Nousek, in particular, for their invaluable help and support. 
This work was supported in Italy by contracts ASI I/088/06/0 and I/023/05/0, at
PSU by NASA contract NAS5-00136. 
We thank P.A.\ Evans and S.\ Vercellone for helpful discussions. 
We also thank the anonymous referee for comments that helped improve the paper.

\bsp

\label{lastpage}


\begin{thebibliography}{23}
\expandafter\ifx\csname natexlab\endcsname\relax\def\natexlab#1{#1}\fi

\bibitem[{{Buccheri} {et~al.}(1983){Buccheri}, {Bennett}, {Bignami}, {Bloemen},
  {Boriakoff}, {Caraveo}, {Hermsen}, {Kanbach}, {Manchester}, {Masnou},
  {Mayer-Hasselwander}, {Ozel}, {Paul}, {Sacco}, {Scarsi}, \&
  {Strong}}]{Buccheri1983}
{Buccheri}, R., {Bennett}, K., {Bignami}, G.~F., {et~al.} 1983, \aap, 128, 245

\bibitem[{{Burrows} {et~al.}(2005){Burrows}, {Hill}, \& {Nousek et
  al.}}]{Burrows2005:XRTmn}
{Burrows}, D.~N., {Hill}, J.~E., \& {Nousek}, J.~A.,  et al., 2005, Space Science
  Reviews, 120, 165

\bibitem[{{Cash}(1979)}]{Cash1979}
{Cash}, W. 1979, \apj, 228, 939

\bibitem[{{Chernyakova} {et~al.}(2003){Chernyakova}, {Lutovinov}, {Capitanio},
  {Lund}, \& {Gehrels}}]{Chernyakova2003}
{Chernyakova}, M., {Lutovinov}, A., {Capitanio}, F., {Lund}, N., \& {Gehrels},
  N. 2003, \ATel, 157

\bibitem[{{Corbet}(1986)}]{Corbet1986}
{Corbet}, R.~H.~D. 1986, \mnras, 220, 1047

\bibitem[{{Davidson} \& {Ostriker}(1973)}]{Davidson1973}
{Davidson}, K. \& {Ostriker}, J.~P. 1973, \apj, 179, 585

\bibitem[{{Ducci} {et~al.}(2009){Ducci}, {Sidoli}, {Mereghetti}, {Paizis}, \&
  {Romano}}]{Ducci2009}
{Ducci}, L., {Sidoli}, L., {Mereghetti}, S., {Paizis}, A., \& {Romano}, P.
  2009, MNRAS, in press, arXiv:0906.3185

\bibitem[{{Giunta} {et~al.}(2009){Giunta}, {Bozzo}, {Bernardini}, {Israel},
  {Stella}, {Falanga}, {Campana}, {Bazzano}, {Dean}, \& {Mendez}}]{Giunta2009}
{Giunta}, A., {Bozzo}, E., {Bernardini}, F., {et~al.} 2009, MNRAS, in press, arXiv:0905.4866

\bibitem[{{Lefever} {et~al.}(2007){Lefever}, {Puls}, \& {Aerts}}]{Lefever2007}
{Lefever}, K., {Puls}, J., \& {Aerts}, C. 2007, \aap, 463, 1093

\bibitem[{{Levine} \& {Corbet}(2006)}]{Levine2006:igr18483}
{Levine}, A.~M. \& {Corbet}, R. 2006, \ATel, 940

\bibitem[{{Negueruela} {et~al.}(2006){Negueruela}, {Smith}, {Reig}, {Chaty}, \&
  {Torrej{\'o}n}}]{Negueruela2005a}
{Negueruela}, I., {Smith}, D.~M., {Reig}, P., {Chaty}, S., \& {Torrej{\'o}n},
  J.~M. 2006, in Proceedings of the ``The X-ray Universe 2005'', 26-30
  September 2005, El Escorial, Madrid, Spain. Ed. by A. Wilson. ESA SP-604,
  Volume 1,  
  165

\bibitem[{{Poole} {et~al.}(2008){Poole}, {Breeveld}, \& {Page et
  al.}}]{Poole2008:UVOTmn}
{Poole}, T.~S., {Breeveld}, A.~A., \& {Page}, M.~J.,  et al., 2008, \mnras, 383, 627

\bibitem[{{Rahoui} \& {Chaty}(2008)}]{Rahoui2008:18483}
{Rahoui}, F. \& {Chaty}, S. 2008, \aap, 492, 163

\bibitem[{{Rappaport} \& {Joss}(1983)}]{Rappaport1983}
{Rappaport}, S.~A. \& {Joss}, P.~C. 1983, in Accretion-Driven Stellar X-ray
  Sources, ed. W.~H.~G. {Lewin} \& E.~P.~J. {van den Heuvel}, 1--39

\bibitem[{{Romano} {et~al.}(2009a){Romano}, {Sidoli}, {Cusumano}, et al.}]{Romano2009:sfxts_paperV}
{Romano}, P., {Sidoli}, L., {Cusumano}, G., {et~al.} 2009a, MNRAS, in press, arXiv:0907.1289

\bibitem[\protect\citeauthoryear{{Romano}, {Sidoli}, {Cusumano}, {Vercellone},
  {Mangano} \& {Krimm}}{{Romano} et~al.}{2009b}]{Romano2009:11215_2008}
{Romano} P.,  {Sidoli} L.,  {Cusumano} G.,  {Vercellone} S.,  {Mangano} V.,
  {Krimm} H.~A.,  2009b, \apj, 696, 2068

\bibitem[{{Roming} {et~al.}(2005){Roming}, {Kennedy}, \& {Mason et
  al.}}]{Roming2005:UVOTmn}
{Roming}, P.~W.~A., {Kennedy}, T.~E., \& {Mason }, K.~O., et al., 2005, Space
  Science Reviews, 120, 95

\bibitem[{{Searle} {et~al.}(2008){Searle}, {Prinja}, {Massa}, \&
  {Ryans}}]{Searle2008}
{Searle}, S.~C., {Prinja}, R.~K., {Massa}, D., \& {Ryans}, R. 2008, \aap, 481,
  777

\bibitem[{{Sguera} {et~al.}(2005){Sguera}, {Barlow}, {Bird}, {Clark}, {Dean},
  {Hill}, {Moran}, {Shaw}, {Willis}, {Bazzano}, {Ubertini}, \&
  {Malizia}}]{Sguera2005}
{Sguera}, V., {Barlow}, E.~J., {Bird}, A.~J., {et~al.} 2005, \aap, 444, 221

\bibitem[{{Sguera} {et~al.}(2006){Sguera}, {Bazzano}, {Bird}, {Dean},
  {Ubertini}, {Barlow}, {Bassani}, {Clark}, {Hill}, {Malizia}, {Molina}, \&
  {Stephen}}]{Sguera2006}
{Sguera}, V., {Bazzano}, A., {Bird}, A.~J., {et~al.} 2006, \apj, 646, 452

\bibitem[{{Sguera} {et~al.}(2007){Sguera}, {Hill}, {Bird}, {Dean}, {Bazzano},
  {Ubertini}, {Masetti}, {Landi}, {Malizia}, {Clark}, \& {Molina}}]{Sguera2007}
{Sguera}, V., {Hill}, A.~B., {Bird}, A.~J., {et~al.} 2007, \aap, 467, 249

\bibitem[\protect\citeauthoryear{{Shimada}, {Ito}, {Hirata} \&
  {Horaguchi}}{{Shimada} et~al.}{1994}]{Shimada1994}
{Shimada} M.~R.,  {Ito} M.,  {Hirata} B.,    {Horaguchi} T.,  1994, in {Balona}
  L.~A.,  {Henrichs} H.~F.,   {Le Contel} J.~M.,  eds, Pulsation; Rotation; and
  Mass Loss in Early-Type Stars Vol.~162 of IAU Symposium, {Radiatively driven
  winds of OB stars}, 487

\bibitem[{{Skinner} {et al.}(2008)}]{Skinner2008}
Skinner, G.,  Tueller, J.,  Beckmann, V.,  Corbet, R., Farrell, S., Krimm, H.A., 
Markwardt, C., 2008, in proceedings of 7th INTEGRAL Workshop,
PoS(Integral08)130 

\bibitem[{{Sidoli}(2009)}]{Sidoli2009:cospar}
{Sidoli}, L. 2009, Advances in Space Research, 43, 1464

\bibitem[{{Sidoli} {et~al.}(2006){Sidoli}, {Paizis}, \&
  {Mereghetti}}]{SidoliPM2006}
{Sidoli}, L., {Paizis}, A., \& {Mereghetti}, S. 2006, \aap, 450, L9

\bibitem[\protect\citeauthoryear{{Sidoli}, {Romano}, {Mangano}, {Pellizzoni},
  {Kennea}, {Cusumano}, {Vercellone}, {Paizis}, {Burrows} \&
  {Gehrels}}{{Sidoli} et~al.}{2008}]{Sidoli2008:sfxts_paperI}
{Sidoli} L.,  {Romano} P.,  {Mangano} V., et al.,  2008, \apj, 687, 1230

\bibitem[{{Swank} {et~al.}(2007){Swank}, {Smith}, \& {Markwardt}}]{Swank2007}
{Swank}, J.~H., {Smith}, D.~M., \& {Markwardt}, C.~B. 2007, \ATel, 999

\end{thebibliography}
\end{document}